\documentclass[pra,twocolumn,superscriptaddress,showpacs]{revtex4-1}

\usepackage{graphicx}
\usepackage{amsmath}
\usepackage{color}
\usepackage[normalem]{ulem} 
\usepackage{hyperref} 

\newcommand{\te}{T_{\rm e}}

\renewcommand{\vec}[1]{\mathbf{#1}}
\renewcommand{\Im}{\mathop{\mathrm{Im}}}
\renewcommand{\Re}{\mathop{\mathrm{Re}}}
\newcommand{\sign}{\mathop{\mathrm{sign}}}%

\newcommand{\rob}[1]{{#1}}

\begin{document}
\title{Local bistability under microwave heating \\ for spatially mapping disordered superconductors}

\author{D. B. Karki}
\affiliation{Division of Quantum State of Matter, Beijing Academy of Quantum Information Sciences, Beijing 100193, China}

\author{R. S. Whitney}
\affiliation{Univ. Grenoble Alpes, CNRS, LPMMC, 38000 Grenoble, France}

\author{D. M. Basko}
\affiliation{Univ. Grenoble Alpes, CNRS, LPMMC, 38000 Grenoble, France}
\begin{abstract}
We theoretically study a strongly disordered superconducting layer heated by near-field microwave radiation from a nanometric metallic tip.
The microwaves heat up the quasiparticles, which cool by phonon emission and conduction away from the heated area.
Due to a bistability with two stable states of the electron temperature under the tip, the heating can be tuned to induce a sub-micron sized normal region bounded by a sharp domain wall between high and low temperature states. We propose this as a local probe to access different physics from existing methods, for example, to map out inhomogeneous superfluid flow in the layer. The bistability-induced domain wall \rob{can} significantly improve its spatial resolution.
\end{abstract}

\date{\today}
\maketitle


\begin{figure}[b]
\includegraphics[width=0.45\textwidth]{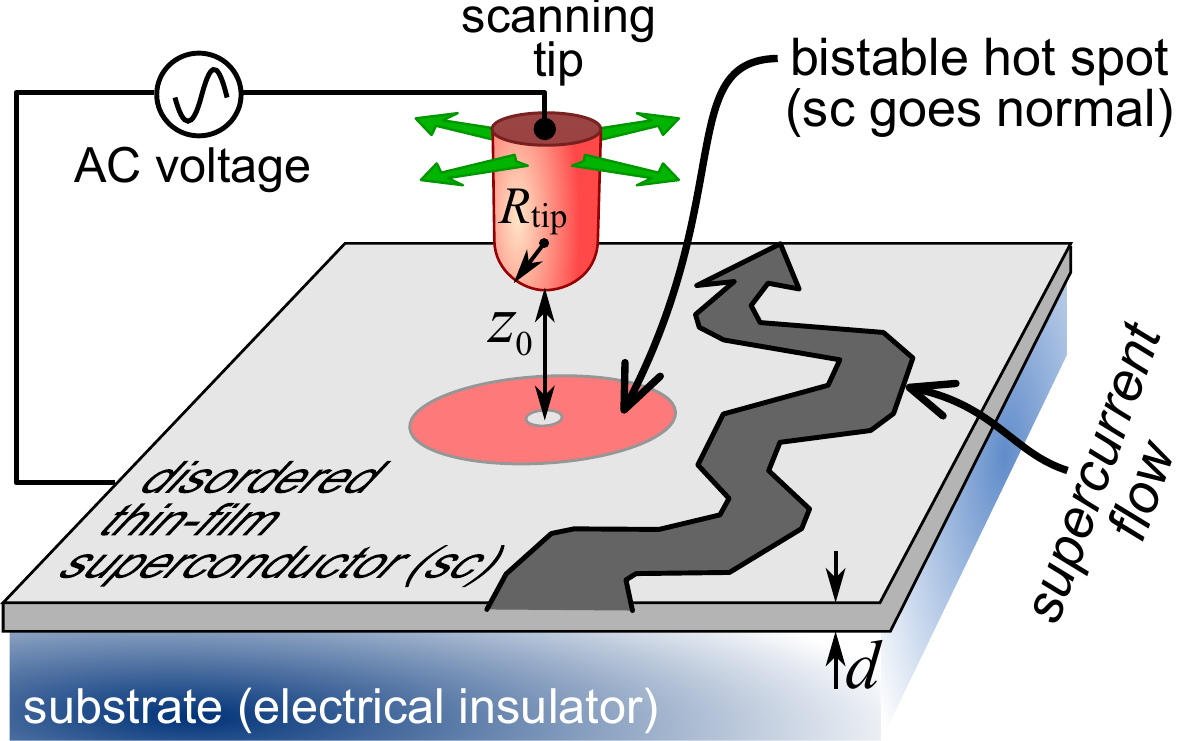}
\caption{Proposed experiment: the ac driven scanning tip creates a hot spot on the surface, where the superconductivity is locally suppressed. 
When the tip passes over places where the supercurrent is flowing, it suppresses this current, thereby providing a map of the supercurrent flow.}
\label{fig_setup}
\end{figure}

\section{Introduction}
Local probes constitute a powerful toolkit for experimental solid state physics. 
Non-invasive local probes of electronic properties range from
the well-known Scanning Tunneling Microscopy (STM) \cite{Tersoff1985Jan,Chen-STM-book} to Microwave Impedance Microscopy (MIM) \cite{Rosner2002, Lai2008}.
Non-invasive local probes of thermal properties have now emerged, such as scanning thermal microscopy~\cite{Gomes2015}, providing a new window into dissipative processes in quantum transport~\cite{Halbertal2016}.
Invasive local probes include scanning gate microscopy (SGM), which probes the spatial structure of inhomogeneous 2D electron gases by measuring changes in global transport properties induced by a local electrostatic perturbation (a charged tip)~\cite{Sellier2011}.  Unfortunately, it cannot probe systems with high electron density (metals or superconductors) because they screen out its electrostatic potential very efficiently.
Here we propose to circumvent this difficulty by applying a local thermal perturbation.

The main idea is to create a local hot spot by applying a microwave drive to a small metallic tip placed near the sample, and to subsequently measure the global transport properties of the sample, see Fig.~\ref{fig_setup}.  Such local heating probe might prove especially suitable for probing thin films of strongly disordered superconductors with short coherence length (a few nanometers), where local heating can create a small normal region. Indeed, while single-electron STM probes the local superconducting gap~\cite{Sacepe2011} and Andreev state microscopy probes the global superconducting phase coherence~\cite{Dubouchet2019}, a local suppression of superconductivity allows one to map out where the supercurrent is flowing in the sample. One can do this by observing if suppressing the superconductivity at a given point induces a significant change in the global supercurrent.

\rob{This spatial mapping of highly disordered superconductors would resolve whether there are spatial regions which are bottlenecks for the supercurrent.  As an application, it would distinguish materials where the supercurrent is approximately uniform, from those in which the superconductivity breaks up into droplets with the supercurrent percolating through only some of them.  
This would clarify when the superconductor-insulator transition is due to a percolation transition, and when it is not.}
This is especially important in view of the potential application of strongly disordered superconductors as superinductance, a key element of superconducting circuit-based quantum technology \cite{Hazard2019,Peruzzo2021}. 
\rob{This probe would also be a thermal analogue of 
the SGM technique, applicable to a variety of nanostructured thin-film materials.}

\rob{The experimental setup would require a low-temperature scanning probe system equipped with microwaves,} similar to the existing MIM  \cite{Rosner2002, Lai2008, Lai2011, Kundhikanjana2011, Allen2019} or near-field scanning microwave microscopy~\cite{Geaney2019} setups.  
\rob{These existing techniques are} intended to operate at low microwave power and leave the sample's properties unmodified, \rob{while we propose a stronger microwave power to modify those properties (locally destroying the superconductivity).} As we will see below, the proposed probe can have higher spatial resolution than the low-temperature scanning laser microscopy~\cite{Zhuravel1996,Sivakov2000}, and can be more versatile due to two independent control parameters, microwave strength and frequency.

The probe's spatial resolution can be improved by a fundamental physics effect: a local overheating bistability for subgap microwave frequencies. It occurs because only quasiparticles get heated by the microwaves, but not the condensate. Hence, while quasiparticles are rare, the superconductor remains cold, and they remain rare. However, once the number of quasiparticles exceeds a threshold, the gap shrinks and the microwave can break the Cooper pairs, generating more quasiparticles, allowing more heating.  This heating is opposed by heat dissipation to phonons, and heat conduction away from the hot spot. This leads to two different (hot and cold) stable steady states in the hot spot, similarly to a global bistability previously known for superconductors subjected to in spatially uniform microwave fields~\cite{Zharov1992, deVisser2010, Thompson2013} or dc currents~\cite{Gurevich1987}.
However, we show a {\it spatially local} bistability can exist so long as the thermal conductivity is not too big.
One consequence of the bistability would be the hysteretic behavior of the electron temperature as the microwave power is turned on or off.
Another consequence, important for the local probe application,
is a sudden domain-wall-like switch from the hotter to the colder steady state at a certain distance from the center of the irradiated region, giving the hot spot a sharp boundary. 
Thus the spatial resolution of the microwave tip is determined by the temperature relaxation length of the sample (the domain wall width), which can be significantly smaller ($\sim20\:\mbox{nm}$) than the hot spot radius, of the order of the tip size ($\sim100\:\mbox{nm}$).

An important assumption behind our arguments is that microwave field can heat only quasipaticles, and that the Cooper pair condensate itself does not have any excitation modes below the superconducting gap~$2\Delta$ which could absorb microwave photons. If such absorption by low-frequency modes is present, this could lead to a significant heating even in the low-temperature state without quasiparticles (such as heating by a low-frequency ac current drive observed in Ref.~\cite{Tamir2019}), which would destabilize the low-temperature state. Then the presence or absence of the local bistability could serve as a probe of the subgap modes of the condensate.

The paper is organized as follows.
In Sec.~\ref{sec:Model} we describe the main ingredients of the model, with some bulky explicit expressions relegated to Appendix~\ref{sec:MattisBardeen} and the derivation of the heating spatial profile given in Appendix~\ref{sec:microwave}.
Sec.~\ref{sec:bistability-local} \rob{treats the simpler} problem of the local bistability in the absence of heat conduction.
Its results are used in Sec.~\ref{sec:bistability-full} to describe the temperature spatial profile in the full problem, with the properties of the domain wall discussed in Appendix~\ref{sec:functional}.
The short Sec.~\ref{sec:thermal} briefly summarizes the results for an alternative heating mechanism: thermal radiation by a hot tip, rather than an applied microwave; the detailed derivations are given in Appendix~\ref{app:thermal}. Finally, in Sec.~\ref{sec:conclusions} we present our conclusions and some remarks regarding potential limitations of our model and possible experimental aspects.

\section{Model}\label{sec:Model}

We consider a tip at a distance $z_0$ above a
strongly disordered superconductor (such as InO$_x$ or NbN) with short coherence length~$\xi$. We assume $\xi$ to be smaller than other length scales of the problem (the tip-plane distance~$z_0$ and the thermal relaxation length~$\Lambda$ defined later), and adopt a model where the superconducting gap~$\Delta$ and the quasiparticle distribution function at each point~$\vec{r}$ correspond to a local equilibrium with a position-dependent electron temperature $\te(\vec{r})$. This implies fast electron-electron collisions and leaves out microwave-induced non-equilibrium effects~\cite{Eliashberg1972, Chang1977, deVisser2014, Semenov2016, Tikhonov2018}.
The phonon temperature $T_\mathrm{ph}$ in the layer is assumed to be fixed by the cryostat, due to a good contact with the substrate.
This simple model, with its standard ingredients, contains only three material parameters: the normal state conductivity~$\sigma_N$, the superconductor critical temperature~$T_c$, and the electron-phonon cooling strength. They all are assumed to be the same as in the bulk material, so the role of the layer thickness~${d}$ is only to relate bulk and surface quantities.

Taking $\te(\vec{r})$ as constant across the layer's thickness,
gives a two-dimensional heat transport equation;
\begin{align}
 c(\te)d\,\frac{\partial\te}{\partial{t}}={}&\boldsymbol{\nabla}\cdot[\mathcal{K}(\te){d}\,\boldsymbol{\nabla}\te]-Q(\te,T_\mathrm{ph}){d}+
 {}\nonumber\\ {}&{}+
 \mathcal{H}(r)\,\frac{I_0^2}{8\pi^2}\Re\frac{1}{\sigma(\omega,\te){d}}.
\label{eq:HeatTransport}
\end{align}
We only consider the stationary state ($\partial\te/\partial t=0$), so the specific heat $c(\te)$ drops out. The bulk electronic thermal conductivity $\mathcal{K}(\te)$ is the textbook expression~\cite{Landau1981,Abrikosov} (see Appendix~\ref{sec:MattisBardeen}); it is $\te\sigma_N/e^2$ multiplied by a function of only $\te/T_c$, and it reduces to the Wiedemann-Franz law for $\te>T_c$.

$Q(\te,T_\mathrm{ph})$ is the power per unit volume, transferred from electrons to phonons. We adopt the standard model of electrons coupled to acoustic phonons~\cite{Chang1977} in which the effective electron-phonon coupling is $\alpha^2(\Omega)\,F(\Omega)\propto\Omega^{n-3}$ for phonon energy~$\Omega$. In particular, $n=5$ when the electron mean free path is much larger than the typical phonon wavelength, and $n=6$ in the opposite limit~\cite{Tsuneto1961,Schmid1973,Reizer1986,Sergeev2000, Catelani2010,Shtyk2013,Savich2017,Vodolazov2017,Nikolic2020}. In the normal state this model yields $Q(\te,T_\mathrm{ph})=\Sigma(\te^n-T_\mathrm{ph}^n)$ with a material-dependent coefficient~$\Sigma$. The expression for $Q(\te,T_\mathrm{ph})$ in the superconductor is rather bulky and given in Appendix~\ref{sec:MattisBardeen}; for each~$n$, $Q(\te,T_\mathrm{ph})$ is given by $\Sigma{T}_c^n$ multiplied by a universal function of $\te/T_c$ and $T_\mathrm{ph}/T_c$. Our results for $n=5$ and $n=6$ are qualitatively similar, as expected~\cite{Catelani2019}; the parameter important for our problem is the differential electron-phonon heat conductance at $\te=T_\mathrm{ph}=T_c$, $\partial{Q}(\te,T_\mathrm{ph})/\partial\te|_{\te=T_\mathrm{ph}=T_c}=n\Sigma{T}_c^{n-1}$. It also defines a crucial length scale in our analysis;
the thermal relaxation length 
$\Lambda\equiv[\mathcal{K}(T_c)/(n\Sigma{T}_c^{n-1})]^{1/2}$.
\begin{table}[]
    \centering
    \begin{tabular}{|c|c|c|c|c|c|}
    \hline
       & ~$T_c$, K~ & ~$1/\sigma_N$, $\Omega\,\mbox{m}$~ & ~$n$~ & ~$\Sigma$, $\mbox{W}\,\mbox{K}^{-n}\,\mbox{m}^{-3}$~ & ~$\Lambda$, nm~ \\
    \hline
    InO$_x$     & 3.5 & $5\times10^{-5}$ & 6 & $2 \times 10^9$ & 17 \\
    \hline
    NbN     & 10.0 & $4\times 10^{-6}$ & 5 & $5 \times 10^9$ & 16 \\
    \hline
    \end{tabular}
    \caption{Material parameters used for the numerical calculations, typical for InO$x$ and NbN \cite{Ovadia2009, Baeva2021}.
    }
    \label{tab:parameters}
\end{table}

The last term of Eq.~(\ref{eq:HeatTransport}) represents Joule heating of the electrons in the layer by the near-field microwaves at frequency~$\omega$. We parametrize its strength by~$I_0$, the amplitude of the total ac displacement current, flowing through the effective capacitor formed by the tip and the layer, due to the applied microwave voltage. The function $\mathcal{H}(r)$ is determined by the spatial distribution of the induced surface currents in the layer; its exact form depends on the tip shape. Still, for any axially symmetric tip whose radius $R_\mathrm{tip}$ does not strongly exceed the tip-sample distance~$z_0$, $\mathcal{H}(r)$ has the same qualitative form: $\mathcal{H}(r=0)=0$, it reaches a maximum value $\sim1/z_0^2$ at $r\sim{z}_0$, and decays at $r\gg{z}_0$. In the calculations we use the expression corresponding to a spherical tip of radius $R_\mathrm{tip}\ll{z}_0$ (see Appendix~\ref{sec:microwave}):
\begin{equation}\label{eq:heating_profile}
    \mathcal{H}(r)=\frac{r^2}{(z_0^2+r^2)(\sqrt{z_0^2+r^2}+z_0)^2}.
\end{equation}
\rob{
This approximation $R_\mathrm{tip}\ll{z}_0$ is not crucial, the shape of $\mathcal{H}(r)$ would have the same qualitative features if one takes $R_\mathrm{tip}\sim {z}_0$, and accounts for the tip not being a sphere. When we estimate the drive strength required for a given heating, we take $R_\mathrm{tip}\sim{z}_0$ to get its order of magnitude.  
}
 
The 2D conductivity $\sigma(\omega,\te){d}$ of the layer also appears in Eq.~(\ref{eq:HeatTransport}), with the bulk conductivity $\sigma(\omega,\te)$ given by the standard Mattis-Bardeen expression~\cite{Mattis1958}, whose real part describes the dissipative response of the quasiparticles, and the imaginary part describes the supercurrent response of the condensate (see Appendix~\ref{sec:MattisBardeen}). For frequencies of the order of the gap and temperatures of the order of~$T_c$ (which will be our focus),  $\Re\sigma$ and $\Im\sigma$ are of the same order as the normal state conductivity. Typical material parameters for two commonly used disordered superconductors, NbN and InO$_x$, are given in Table~\ref{tab:parameters}. 

\section{Bistability in absence of heat conduction}
\label{sec:bistability-local}
It is instructive to start with the local version of Eq.~(\ref{eq:HeatTransport}), setting $\mathcal{K}=0$. Then, at each point $\vec{r}$, the electron temperature $\te$ is found from the algebraic equation
\begin{equation}\label{eq:HeatBalance}
    j^2\Re\frac1{\sigma(\omega,\te)} = Q(\te,T_\mathrm{ph}),
\end{equation}
with $j^2=I_0^2\mathcal{H}(r)/(8\pi^2d^2)$. Equation~(\ref{eq:HeatBalance}) contains only bulk quantities, and is analogous to the heat balance equation for a superconductor in a spatially uniform microwave field. For that problem, the heat balance equation may have two stable solutions for $\te$ \cite{Zharov1992, deVisser2010, Thompson2013}.

\begin{figure}
\includegraphics[width=0.45\textwidth]{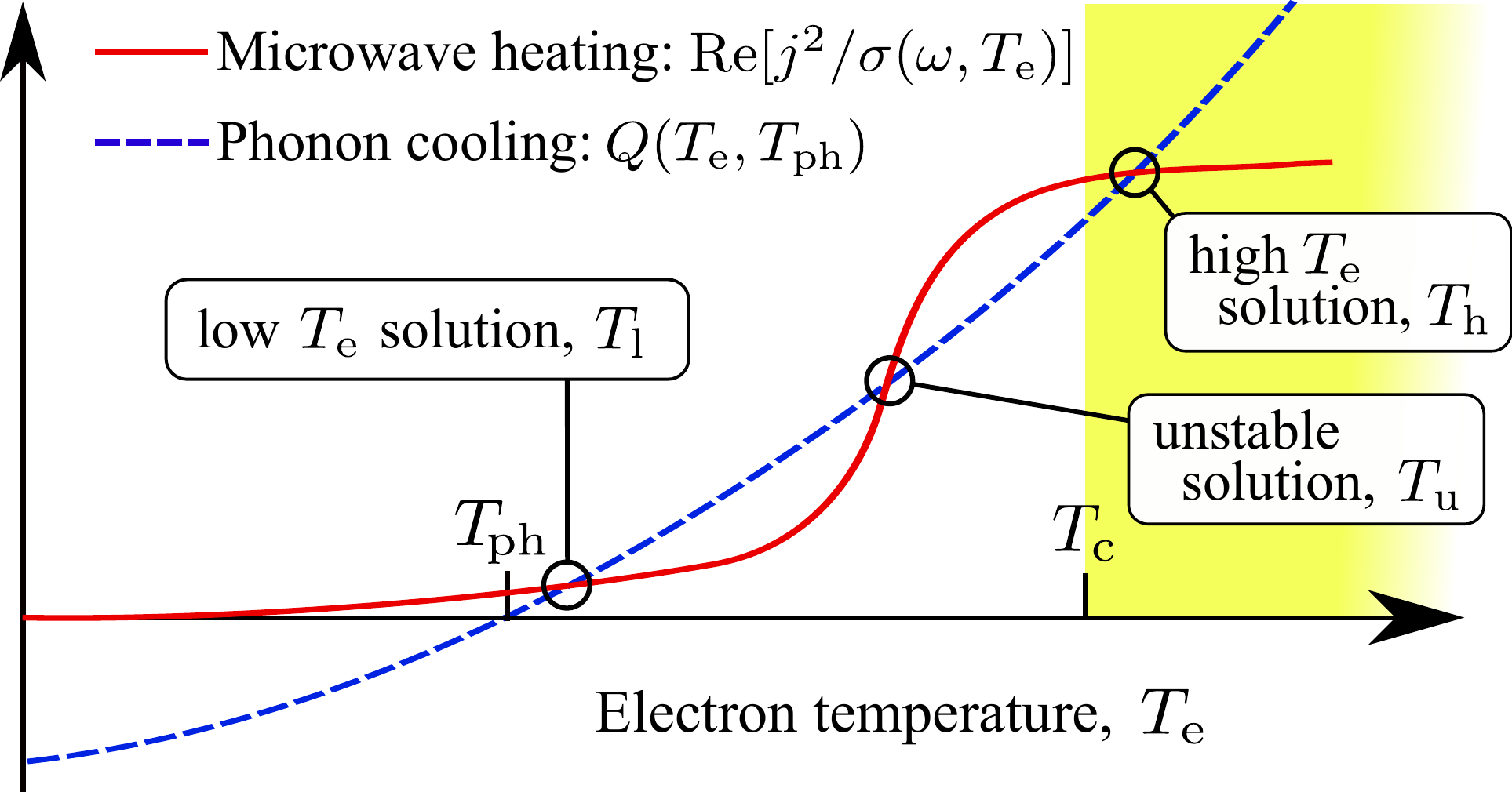}
\caption{A sketch of the $\te$ dependence of the heating and cooling power, the two sides of Eq.~(\ref{eq:HeatBalance}) (red solid and blue dashed curves, respectively). The crossings between the two curves represent the multiple solutions of Eq.~(\ref{eq:HeatBalance}) for~$\te$. Solutions in the yellow region correspond to the normal state.
}
\label{fig_illustrate_bistability}
\end{figure}

The peculiar $\te$ dependence of the heating term on the left-hand side of Eq.~(\ref{eq:HeatBalance}), sketched in Fig.~\ref{fig_illustrate_bistability}, is a rather common origin for bistabilities related to electron overheating~\cite{Gurevich1987}. 
In our case, it is due to the physics of quasiparticle heating, summarized in the Introduction.
As a result, Eq.~(\ref{eq:HeatBalance}) has three solutions for $\te$, as sketched in Fig.~\ref{fig_illustrate_bistability}; only the high- and the low-temperature solutions are stable, the middle one is unstable. Fig.~\ref{fig_illustrate_bistability} shows that $j^2$ controls the vertical scale of the heating curve, so the bistability disappears if $j^2$ is too large or too small.

When $\hbar\omega$ and the solutions for $\te$ are all of the order of $T_c$, the typical scale of $j$ which governs the bistability, is ${j}_*=\sqrt{\sigma_N\Sigma{T}_c^n}$. It is the current density needed to maintain the electrons at $\te=T_c$ when the phonons are at $T_\mathrm{ph}=0$. It is important that $j_*\ll{j}_c$ for the critical current density ${j}_c$; this condition is necessary to justify the calculation of the dissipated power using the linear response conductivity.
(When non-linear effects become noticeable, the physics becomes much more rich~\cite{Baryshev2007}, going beyond the scope of the present study).
The condition $j_*\ll{j}_c$ is naturally satisfied when electron-phonon coupling is weak. Indeed, we obtain $j_*/j_c\sim\sqrt{\hbar/(T_c\tau_\mathrm{ph})}$, 
using (i) the expression $j_c\approx1.5\,\sigma_N[\Delta_0/(2e)](\hbar{D}/\Delta_0)^{-1/2}$ \cite{Annunziata2010} with $2\Delta_0\approx3.53\,T_c$ being the zero-temperature gap and $D$~the electron diffusion coefficient, 
(ii) the Einstein relation $\sigma_N=2N_0e^2D$ with $N_0$ being the density of states per spin at the Fermi level, and 
(iii) that $\Sigma{T}_c^n\sim{N}_0T_c^2/\tau_\mathrm{ph}$ where $\tau_\mathrm{ph}$ is the time the electron with energy $\sim{T}_c$ spends before emitting a phonon. The ratio $j_*/j_c$ must necessarily be small in any material well-described as a gas of electronic quasiparticles.

\begin{figure}
\includegraphics[width=0.48\textwidth]{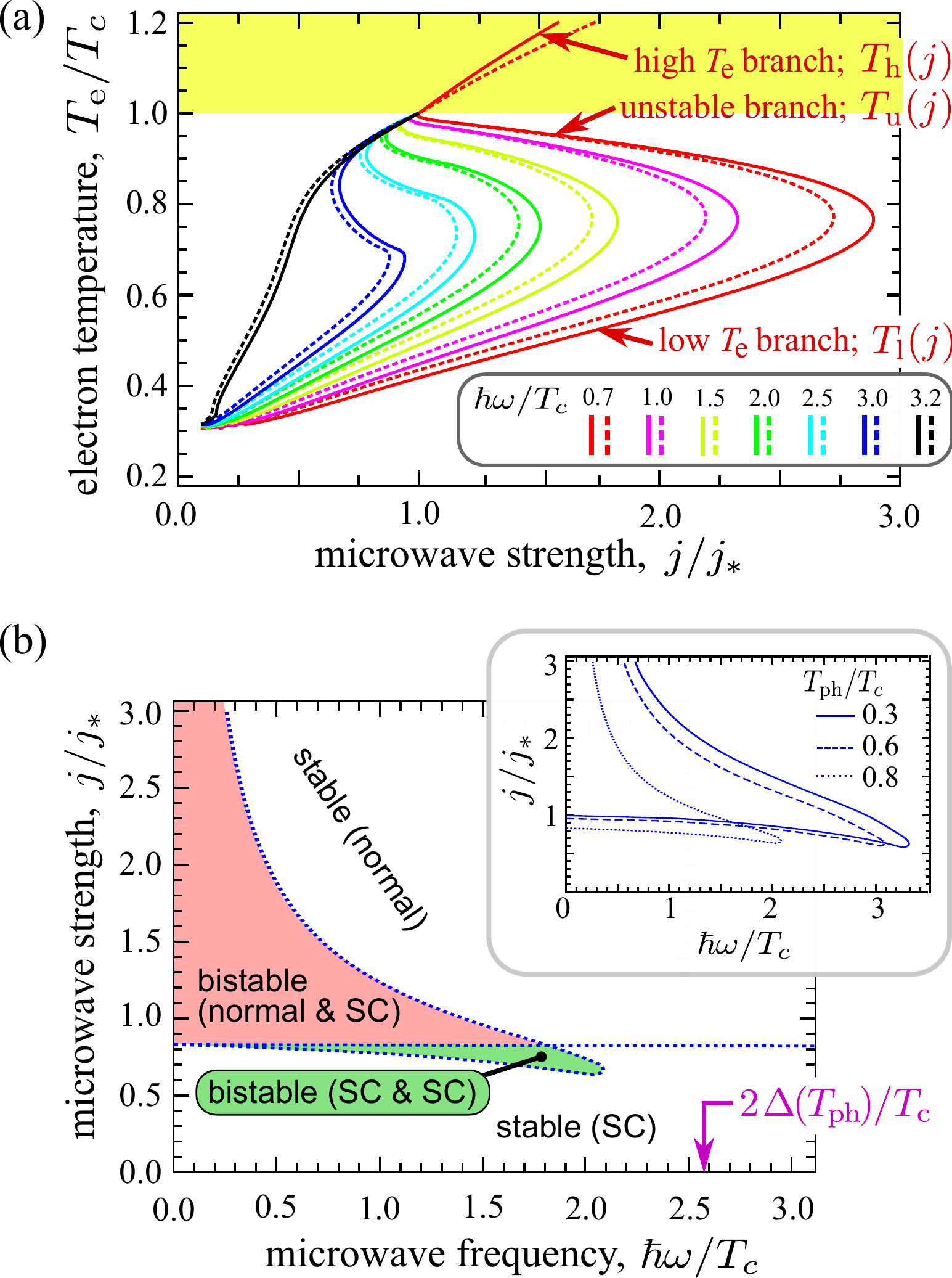}
\caption{Bistability under a spatially uniform microwave drive. (a)~$\te$~as a function of the driving current $j$ for different microwave frequencies $\omega$ and $T_\mathrm{ph}/T_c=0.3$. 
The curves can have three branches (indicated for the red solid curve)
corresponding to three solutions, $T_\mathrm{l}(j)<T_\mathrm{u}(j)<T_\mathrm{h}(j)$, 
the middle one being unstabkle.
Solutions in the yellow region correspond to the superconductor going normal.
Solid and dashed curves correspond to electron-phonon cooling exponents $n=5$ and $n=6$, respectively.  
(b)~The phase diagram in the $(\omega,j)$ plane for $n=5$ and $T_\mathrm{ph}/T_c=0.8$. The white regions have one stable state, superconducting (SC) or normal, as indicated. The pink/green regions have two stable states of the types indicated (there is no region where both states are normal). Inset: the same plot for $T_\mathrm{ph}/T_c=0.3,\,0.6,\,0.8$ (solid, dashed, and dotted curves, respectively).
}
\label{figure_uniform_bist}
\end{figure}

Fig.~\ref{figure_uniform_bist} shows the results of numerical solutions of Eq.~(\ref{eq:HeatBalance}).  The curves are universal when plotted in the appropriate units ($T_c$ and $j_*$), i.e., valid for any material with the electron-phonon cooling exponent $n=5$ or $n=6$. We see that they are very weakly sensitive to whether $n=5$ or $n=6$. The bistable region exists only for frequencies below $2\Delta(T_\mathrm{ph})$, the gap at~$T_\mathrm{ph}$. At low frequencies, (i)~it is bounded from below by the current $j_*\sqrt{1-(T_\mathrm{ph}/T_c)^n}$, (ii)~it extends to high currents $\propto1/\omega^2$ since $\Re[1/\sigma(\omega,\te)]\propto\omega^2$ is small. Of course, at high currents the validity of the theory is limited by the condition $j\ll{j}_c$, which is not included in the model. The high-temperature solution lies below $T_c$ only in a small part of the bistable region with $j<j_*\sqrt{1-(T_\mathrm{ph}/T_c)^n}$; at higher~$j$, the high-temperature solution is normal.

\section{Bistability in the full problem}
\label{sec:bistability-full}
Returning to the steady state of the full Eq.~(\ref{eq:HeatTransport}), we consider 
$j=j(r)=I_0\sqrt{\mathcal{H}(r)/(8\pi^2d^2)}$ 
which vanishes at $r\to0,\infty$, and reaches a maximum $j_\mathrm{max}\approx0.120\,I_0/(z_0d)$ at $r_\mathrm{max}\approx1.27\,z_0$.
Then, for given $\hbar\omega/T_c$ and $T_\mathrm{ph}/T_c$, the solutions are determined by two dimensionless parameters, $j_\mathrm{max}/j_*$ and $\Lambda/z_0$.
Dividing Eq.~(\ref{eq:HeatTransport}) by $nd\Sigma{T}_c^n$ and measuring $r$ in the units of $z_0$, we cast the last two terms of Eq.~(\ref{eq:HeatTransport}) [those entering Eq.~(\ref{eq:HeatBalance})] in the dimensionless parameters ($\te/T_c$ and $j/j_*$) at each~$r$, while the gradient term is proportional to $(\Lambda/z_0)^2$.  From the values in Table~\ref{tab:parameters} we see that $\Lambda\sim20\:\mbox{nm}$ can be significantly smaller than $z_0\sim100\:\mbox{nm}$.

The total power, needed to maintain $j_\mathrm{max}=j_*$, is given by the spatial integral of the last term in Eq.~(\ref{eq:HeatTransport}), $5.5\,\Sigma{T}_c^nz_0^2d\,\ln[(4\pi\sigma_N/\omega)(d/z_0)]$ (see Appendix~\ref{sec:microwave}). This gives a few nanowatts for InO$_x$ and a few microwatts for NbN of thickness $d=10\:\mbox{nm}$ and $z_0=100\:\mbox{nm}$.

\begin{figure}
\includegraphics[width=0.48\textwidth]{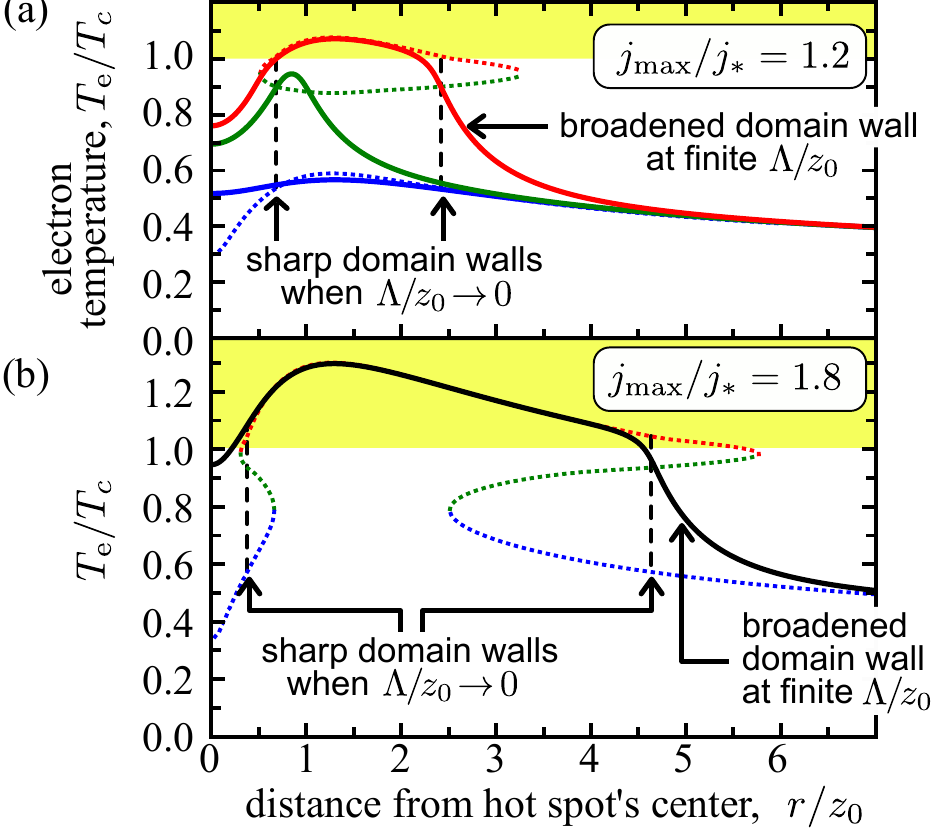}
\caption{Solid curves are numerical solutions of Eq.~(\ref{eq:HeatTransport}) for $\Lambda/z_0=0.16$,  $n=5$, $T_\mathrm{ph}/T_c=0.3$, $\hbar\omega/T_c=2$.  $j_\mathrm{max}/j_*=1.2$ (a)  or 1.8 (b). 
The superconductor goes normal whenever these curves are in the yellow region.
In (a) there are three solutions; high $T_\mathrm{e}$ (red), unstable (green), and low $T_\mathrm{e}$ (blue), in (b) there is only one solution (black).
Dotted curves are the $\Lambda/z_0=0$ solutions of Eq.~(\ref{eq:HeatBalance}) at each $r$.  Domain walls between the low- and the high-$T_\mathrm{e}$ solutions for $\Lambda/z_0\to 0$ (vertical lines ) are given by Eq.~(\ref{eq:domain_wall}), but are broadened for $\Lambda/z_0=0.16$.
}
\label{fig_solutions}
\end{figure}

For $\Lambda/z_0\to0$, the stationary profile $\te(r)$ is determined by the simple mapping of $j(r)/j_*$ to the $j/j_*$ axes on Fig.~\ref{figure_uniform_bist}.
Let $\hbar\omega/T_c$ and $T_\mathrm{ph}/T_c$ be such that the uniform system is bistable in the interval $j_1<j<j_2$ with some $j_1,j_2$; that is, Eq.~(\ref{eq:HeatBalance}) has three solutions $T_\mathrm{l}(j)<T_\mathrm{u}(j)<T_\mathrm{h}(j)$ (see Fig.~\ref{figure_uniform_bist}a) for $j_1<j<j_2$, only $T_\mathrm{l}(j)$ for $j<j_1$, and only $T_\mathrm{h}(j)$ for $j>j_2$.
Thus, for $j_\mathrm{max}<j_1$, only the solution $T_\mathrm{l}(j(r))$ is possible for any~$r$.
For $j_1<j_\mathrm{max}<j_2$, there are three local solutions $T_\mathrm{l}\big(j(r)\big)<T_\mathrm{u}\big(j(r)\big)<T_\mathrm{h}\big(j(r)\big)$ in an interval of $r$ around $r_\mathrm{max}$ where $j(r)>j_1$, so the dependence $\te(r)$ consists of a low-temperature branch and a disconnected closed contour [dotted curve in Fig.~\ref{fig_solutions}(a)]. 

For $j_\mathrm{max}>j_2$, only the $T_\mathrm{h}(j(r))$ solution is possible around $r_\mathrm{max}$, where $j(r)>j_2$, but there are two regions with three solutions on both sides [dotted curve in Fig.~\ref{fig_solutions}(b)]. 
Thus the system must switch between the stable low- and high-temperature local solutions creating a ``domain wall'' (whose width is vanishing in the limit $\Lambda/{z}_0\to0$) at some position~$r$ on each side of $r_\mathrm{max}$. This position can be found by noting that any global stable stationary solution of Eq.~(\ref{eq:HeatTransport}) represents a minimum of a certain functional $\mathcal{F}[\te(\vec{r})]$, explicitly given in Appendix~\ref{sec:functional}. 
Minimizing the functional with respect to the domain wall position~$R$, we identify it as the point where $j(R)$ satisfies the condition
\begin{align}
&\int_{T_\mathrm{l}(j(R))}^{T_\mathrm{u}(j(R))}\left[Q(T,T_\mathrm{ph})-\Re\frac{j^2(R)}{\sigma(\omega,T)}\right]\mathcal{K}(T)\,dT={}\nonumber\\
&{}=\int_{T_\mathrm{u}(j(R))}^{T_\mathrm{h}(j(R))}\left[\Re\frac{j^2(R)}{\sigma(\omega,T)}
-Q(T,T_\mathrm{ph})\right]\mathcal{K}(T)\,dT,
\label{eq:domain_wall}
\end{align}
which resembles the Maxwell construction (equal-area rule) for the Van der Waals isotherms,
as was also noted for overheated superconductors subject to dc currents~\cite{Gurevich1987}.
This construction is valid for $\Lambda/z_0\to 0$, but it remains qualitatively correct for realistic values of $\Lambda/z_0$, as we will see below.

Then for $j_\mathrm{max}>j_2$, the system has only one global solution for $T_\mathrm{e}(r)$. However, it is the underlying local bistability that causes the domain walls in that solution, whose width $\Lambda$ can be significantly smaller than the size of the hot spot (solid black curve in Fig.~\ref{fig_solutions}b). This rapid spatial variation of $\te(\vec{r})$ will be useful for creating a local thermal perturbation of the system with sub-micron resolution. Intriguingly, $\te$ exceeds $T_c$ in a ring, leaving a small superconducting core precisely below the tip.

In contrast, for $j_1<j_\mathrm{max}<j_2$ the disconnected low-temperature branch $T_\mathrm{l}(r)$ represents a stable global solution of Eq.~(\ref{eq:HeatTransport}) in the limit $\Lambda/z_0\to0$. In addition, there is another stable solution with two domain walls. Then, the system exhibits a global bistability [multiple solutions of Eq.~(\ref{eq:HeatTransport}) for the whole profile~$\te(r)$], inherited from the local bistability (multiple solutions of Eq.~(\ref{eq:HeatBalance}) for $\te$ at a given point~$r$).
The global bistability is accompanied by a hysteretic behaviour of $\te(r)$ as one changes the microwave strength, $j_\mathrm{max}/j_*$.

\begin{figure}
\includegraphics[width=0.48\textwidth]{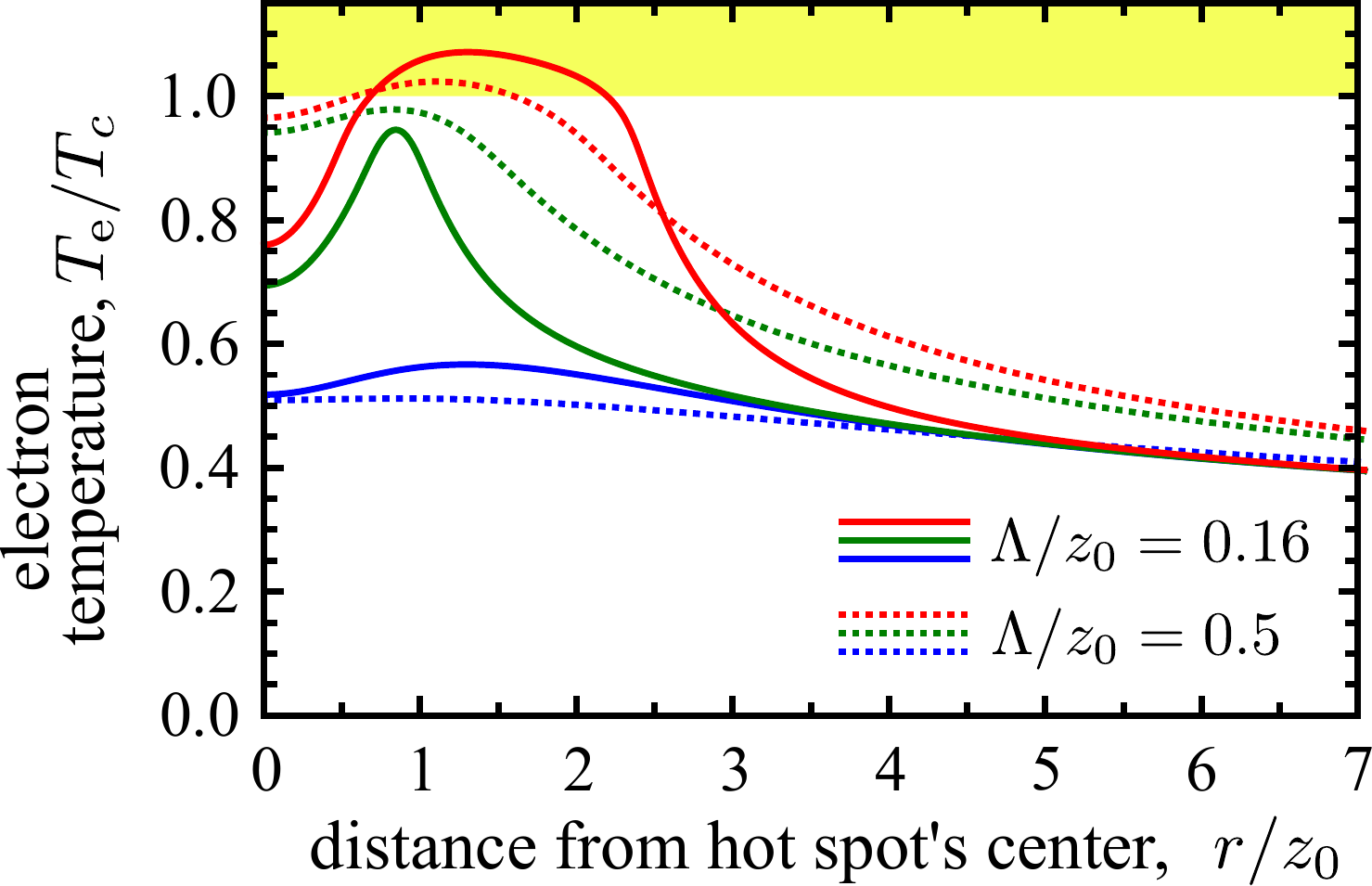}
\caption{Solid curves the same as in Fig.~\ref{fig_solutions}a, and
dotted curves have larger $\Lambda/z_0$. 
The high $T_\mathrm{e}$ (red) and unstable (green) solutions
approach each other as $\Lambda/z_0$ increases, so they meet and annihilate at a given $\Lambda/z_0$ (here  $\Lambda/z_0\approx0.6$), leaving only the low $T_\mathrm{e}$ (blue) solution at higher $\Lambda/z_0$.}
\label{fig_solutions_z0}
\end{figure}

Finite $\Lambda/z_0$ causes a broadening of the domain walls; the solid curves
in Fig.~\ref{fig_solutions} show this for realistic parameters.
Fig.~\ref{fig_solutions_z0} shows that increasing $\Lambda/z_0$ causes the high-temperature and unstable solutions to approach each other, and annihilate at a certain $\Lambda/z_0$ (e.g., $\Lambda/z_0\approx0.6$ for the parameters in Fig.~\ref{fig_solutions_z0}). The low-temperature solution survives at larger $\Lambda/z_0$. because it is favoured by the faster heat evacuation from the below the tip.

\section{Local heating with a hot tip}
\label{sec:thermal}
%
Finally, we briefly discuss another possible setup, when the sample is heated not by the external microwave drive, but by thermal radiation from the tip, held at high temperature~$T_\mathrm{tip}$. The corresponding microwave field can be modelled by that of a thermally fluctuating electric or magnetic dipole, as discussed in detail in Appendix~\ref{app:thermal}. The heating power is then given by an integral over all frequencies. Crucially, both the typical frequency and the strength of the microwave field are controlled by the same parameter $T_\mathrm{tip}$, while in the previous setting the microwave strength $I_0$ and frequency~$\omega$ could be controlled independently. For typical material parameters, to produce a noticeable change in~$\te$, one needs $T_\mathrm{tip}\gg{T}_c$. Then the heating is due to absorption of photons with $\omega\gg{T}_c/\hbar$, so it is not sensitive to the superconductivity. This results in a single solution, a smooth profile $\te(r)$ exceeding $T_\mathrm{ph}$ in a region whose size is a few~$z_0$ at least. 
Moreover, for realistic parameters,  its effect is too weak to locally destroy the superconductivity (unless $T_\mathrm{ph}\to{T}_c$), so it would be a much worse local probe than a tip with microwave driving.

\section{Concluding remarks}
\label{sec:conclusions}
We propose a local thermal probe based on a submicron-sized hot spot created in a thin superconducting layer by microwave radiation produced by a small metallic tip. Our simple model shows how the electron temperature is locally driven away from the substrate (phonon) temperature, assuming the electrons remain in local thermal equilibrium among themselves. We have shown that the hot spot can have two possible stable states, similarly to a bulk bistability, discussed earlier for spatially uniform microwave fields. 

We have identified the superconductor's thermal relaxation length $\Lambda$, 
and shown that global bistability requires $\Lambda\lesssim{z}_0$, 
the tip-sample distance (or the tip size, if larger). 
This is the case for strongly disordered superconductors such as NbN or InO$_x$. Then the hot spot has a sharp boundary corresponding to a domain wall between two local stable solutions. The hot spot temperature can be tuned to locally destroy the superconductivity.  We thus propose it as a scanning probe with sub-micron resolution, ideal for mapping out where the supercurrent flows in such disordered superconductors.

The proposed probe has an advantage of having two control parameters (the microwave's frequency and its strength). The requirements on the geometry are not very stringent: it is only important that the tip size and the tip-sample separation are not too large (50--100~nm). Since the dependence of the field on the tip-sample distance is not exponential, maintaining a constant tip-sample distance is not so crucial as, e.~g., for STM. In most cases, a few percent change in $z_0$ will lead to a few percent change in the temperature or in the domain wall position, unless one hits the point where a stable solution disappears.

The model we used in our calculations was developed \rob{for} homogeneously disordered superconductors, but the qualitative conclusions may have wider applicability. Our calculations were based on the assumption of position-independent local conductivity, which always breaks down below some length scale. We had in mind the simple situation when this scale is given by the superconducting coherence length, but it can be something else.  For granular systems (such as granular aluminum, a promising material for superconducting circuits~\cite{Grunhaupt2019, Winkel2020}) one can describe the system by a local conductivity on length scales longer than the typical grain size. If the grain size happens to be larger than the thermal relaxation length, then it would be the \rob{grain size} that determines the domain wall size.

In a disordered superconductor the superconducting gap may fluctuate in space, typically on the scale of the coherence length. Since the spatial scale of the absorption profile in Eq.~(\ref{eq:heating_profile}) is assumed to be much larger than the coherence length (the basic assumption behind our local approach), one can effectively replace the position-dependent $\Re[1/\sigma(\omega,\te)]$ in Eq.~(\ref{eq:HeatTransport}) by its spatial average. This average will have a smoother dependence on $\te$ than that at a fixed gap, which would lead to shrinking of the bistable region in Fig.~\ref{figure_uniform_bist}(b). However, to completely kill the bistability, the disorder must be sufficiently strong to introduce microwave absorption at low frequencies, that is, gap fluctuations must be of the order of the gap itself. Such strong fluctuations \rob{may} occur in some rare regions \rob{of the disordered superconductor}; the proposed probe could then serve to identify such regions.

\acknowledgments
We thank B. Sac\'ep\'e, H. Sellier, and K. Tanigaki for illuminating discussions. This work is supported by the project TQT (ANR-20-CE30-0028) of the French
National Research Agency (ANR).

\appendix

\section{Electric and thermal conductivity, electron-phonon cooling}
\label{sec:MattisBardeen}

Here, we briefly summarize how the  various parameters in Eq.~(\ref{eq:HeatTransport}) are modeled.

For the temperature dependence of the gap $\Delta(\te)$ we use an empirical expression~\cite{Sheahen1966},
\begin{equation}
\frac{\Delta(\te)}{\Delta_0}=\sqrt{\cos\left(\frac{\pi}{2}\,\frac{\te^2}{T_c^2}\right)},
\end{equation}
which agrees within 3\% with the BCS expression. $\Delta_0$~is related to~$T_c$ by the weak coupling relation ($\gamma=0.577\ldots$ is the Euler-Mascheroni constant):
\begin{equation}
\frac{\Delta_0}{T_c}=\frac{\pi}{e^{\gamma}}\approx1.76.
\end{equation}

The ac conductivity of a superconductor is~\cite{Mattis1958}
\begin{align}
 \frac{\sigma(\omega)}{\sigma_N}={}&{} \frac{1}{\hbar\omega}\int\limits^\infty_{\Delta} d\epsilon\,\Upsilon(\epsilon)\left(\tanh\frac{\epsilon+\hbar\omega}{2\te}-\tanh\frac{\epsilon}{2\te}\right){}\nonumber\\
&{}-\frac{\Theta(\hbar\omega-2\Delta)}{\hbar\omega}\int\limits^{-\Delta}_{\Delta-\hbar\omega} d\epsilon\,\Upsilon(\epsilon)\tanh\frac{\epsilon+\hbar\omega}{2\te} \nonumber\\
&{}-\frac{i}{\hbar\omega}\int\limits^\Delta_{\max\{\Delta-\hbar\omega, -\Delta\}}d\epsilon\,\Upsilon(\epsilon)\tanh\frac{\epsilon+\hbar\omega}{2\te},
\label{eq:MattisBardeen}
\end{align}
where
\begin{equation}
\Upsilon(\epsilon)=\frac{\epsilon(\epsilon+\hbar\omega)+\Delta^2}{\sqrt{|\epsilon^2-\Delta^2|}\sqrt{(\epsilon+\hbar\omega)^2-\Delta^2}}.
\end{equation}
Here $\sigma_N$ is the bulk electrical conductivity of the material in the normal state. The electronic contribution to the thermal conductivity of the superconductor also depends on $\sigma_N$ via~\cite{Landau1981,Abrikosov}:
\begin{equation}
\mathcal{K}(\te)=\frac{\sigma_N}{e^2}\int^\infty_{\Delta(\te)}\frac{\epsilon^2\,d\epsilon}{2\te^2\cosh^2[\epsilon/(2\te)]}.
\end{equation}

The power per unit volume transferred from electrons to phonons can be written using the collision integral in Ref.~\cite{Chang1977} as
\begin{align}
    Q(\te,T_\mathrm{ph})={}&{}
    \frac\Sigma{4(n-1)!\,\zeta(n)}\int_{-\infty}^\infty{d}\epsilon
    \int_0^\infty{d}\Omega\,\Omega^{n-2}
    \times{}\nonumber \\ {}&{}\times
    \left(\coth\frac\Omega{2\te}-\coth\frac\Omega{2T_\mathrm{ph}}\right)
    \times{}\nonumber \\ {}&{}\times
    \left(\tanh\frac{\epsilon+\Omega}{2\te}-\tanh\frac\epsilon{2\te}\right)
    \times{}\nonumber \\ {}&{}\times
    \frac{\Theta(|\epsilon|-\Delta)}{\sqrt{\epsilon^2-\Delta^2}}\,
\frac{\Theta(|\epsilon+\Omega|-\Delta)}{\sqrt{(\epsilon+\Omega)^2-\Delta^2}}
    \times{}\nonumber \\ {}&{}\times
\left[\epsilon(\epsilon+\Omega)-\Delta^2\right]\sign[\epsilon(\epsilon+\Omega)].
\end{align}

\section{Heating by a microwave drive}
\label{sec:microwave}

We model the superconducting layer as an infinitely thin 2D sheet with a 2D conductivity $\sigma_{2D}(\omega)=\sigma(\omega){d}$, given by the standard Mattis-Bardeen expression~(\ref{eq:MattisBardeen}) multiplied by the layer thickness~${d}$. 
We represent the tip as a small spherical particle of radius~$R_\mathrm{tip}$, placed at a distance $z_0$ from the  layer, assuming both ${d},R_\mathrm{tip}\ll{z}_0$. While the first inequality is quite realistic (we have in mind $z_0\sim100\:\mbox{nm}$, ${d}\sim10-20\:\mbox{nm}$), the second will be used to simplify the calculations and obtain relatively simple final expressions; in the end we will set $R_\mathrm{tip}\sim{z}_0$, so these expressions will be valid only as qualitative estimates (which would be the case anyway, since in reality the tip is not spherical). This geometry determines the tip-layer capacitance 
$C_\mathrm{tip}\sim4\pi\varepsilon_0{R}_\mathrm{tip}$, where $\varepsilon_0$ is the vacuum dielectric permittivity.

If a microwave voltage is applied between the sample and the tip, this situation can be analyzed in terms of an effective circuit which includes, in series with the microwave voltage source at frequency~$\omega$, and the tip-sample capacitance $C_\mathrm{tip}$, an effective impedance $Z_\mathrm{s}(\omega)$ representing the sample, as well as an external impedance $Z_\mathrm{ext}$ corresponding to the external circuit used to connect the voltage source.

Typically, the impedance of the capacitor, $Z_\mathrm{tip}=-1/(i\omega{C}_\mathrm{tip})$, is much larger than the sample impedance~$Z_\mathrm{s}$. Indeed, associating the latter with the resistance per square $1/\sigma_{2D}$ of the 2D layer with thickness ${d}\ll{z}_0$, estimating the capacitance to be of the order of the tip radius, and taking the latter to be $50\:\mbox{nm}$, for  $\hbar\omega=1\:\mbox{K}$ we obtain $Z_\mathrm{tip}=1.4\:\mbox{M}\Omega$. This is much larger than $1/\sigma_{2D}$ even for $\sigma_{2D}=4e^2/(2\pi\hbar)=1/(6.46\:\mbox{k}\Omega)$,
about the smallest possible sheet conductivity allowed for a superconductor, below which a superconductor-insulator transition occurs~\cite{Jaeger1986, Haviland1989}. This means that the oscillating charge distribution in the layer is mainly determined by the oscillating charge $q(t)$ on the tip, so the currents in the layer follow to maintain this oscillating charge distribution, and the small in-plane electric field is the one required to drive these currents.

The external impedance $Z_\mathrm{ext}$ depends on the specific experimental setup. If $Z_\mathrm{ext}\gg{Z}_\mathrm{tip}$, the tip is effectively current-biased. An applied current $I(t)=I_0\cos\omega{t}=dq/dt$ corresponds to the charge on the tip $q(t)=(I_0/\omega)\sin\omega{t}$. If $Z_\mathrm{ext}\ll{Z}_\mathrm{tip}$, the tip is biased by a voltage $V(t)=V_0\cos\omega{t}$, so the charge on the tip is $q(t)=C_\mathrm{tip}V_0\cos\omega{t}$.

In electrostatics, a point charge $q$ placed at a distance $z_0$ from a conducting plane, induces a 2D screening charge density on the plane~\cite{JacksonBook}
\begin{equation}\label{eq:imagecharge}
\rho(\vec{r})  =  -\frac{q}{2\pi}\frac{z_0}{(r^2+z_0^2)^{3/2}}.
\end{equation}
where $\vec{r}=(x,y)$ is the in-plane position, and $r=|\vec{r}|$.
When the tip charge $q(t)$ is oscillating, we assume that the density $\rho(\vec{r})$ simply follows Eq.~(\ref{eq:imagecharge}) instantaneously. This instantaneous approximation breaks down at sufficiently large distances where the charges are no longer able to follow. The charge density relaxation time at a distance~$r$ can be estimated as the $RC$ relaxation time of an effective circuit with the capacitance $\sim4\pi\varepsilon_0{r}$ and the resistance $\sim1/\sigma_{2D}$ (since in 2D the resistance per square does not scale with the size). Requiring this relaxation time to be smaller than $1/\omega$, we arrive at the length scale $a_\mathrm{e}(\omega)=i\sigma_{2D}/(2\varepsilon_0\omega)$ (see Eq.~(\ref{eq:phielectricdipole}) for the rigorous definition). Another obvious cutoff scale is $c/\omega$, which becomes more relevant if $\sigma_{2D}/(2\varepsilon_0c)>1$. For the frequencies we are interested in (of the order of the superconductor critical temperature, a few Kelvins), both cutoff scales are much larger than~$z_0$, the typical length scale of Eq.~(\ref{eq:imagecharge}).
Thus they play almost no role in the physics discussed in this work, they only enter via a logarithmic cutoff in Eq.~(\ref{Eq:total_power}) below.

The 2D current density $\vec{J}(\vec{r},t)=\vec{j}(\vec{r},t){d}$ can be found from the continuity equation $\boldsymbol{\nabla}\cdot\vec{J} = -\partial\rho/\partial{t}$ with $\rho(\vec{r},t)$ given by Eq.~(\ref{eq:imagecharge}). Since the whole picture is axially symmetric, $\vec{J}$ has only the radial component $J_r$ which satisfies a first-order ordinary differential equation with $I=dq/dt$:
\begin{equation}
\frac{\partial {J_r}}{\partial r}+\frac{J_r}{r}-\frac{I} {2\pi} \frac{z_0}{(r^2+z_0^2)^{3/2}} =0.
\end{equation}
The solution of this equation which is finite at $r\to0$, reads
\begin{equation}
J_r(r,t)=\frac{I(t)}{2\pi{r}}\left(1-\frac{z_0}{\sqrt{z_0^2+r^2} }\right).
\end{equation}
For the current $I(t)=I_0\cos\omega{t}$, the in-plane electric field is related to this current via $\sigma_{2D}(\omega)$, so the Joule dissipation per unit area is given by
\begin{equation}
    P(r)=\frac{1}{(2\pi{r})^2}\left(1-\frac{z_0}{\sqrt{z_0^2+r^2} }\right)^2\frac{I_0^2}2\Re\frac1{\sigma_{2D}(\omega)}.
    \label{Eq:P(r)}
\end{equation}
The total power injected into the sample is the integral of $P(r)$ over the sample area,
\begin{equation}
    \int_0^\infty{P}(r)\,2\pi{r}\,dr=\frac{I_0^2}{4\pi}\Re\frac1{\sigma_{2D}(\omega)}\ln\frac{\min\{c/\omega,|a_\mathrm{e}(\omega)|\}}{z_0},
    \label{Eq:total_power}
\end{equation}
where the logarithmic divergence at large distances is cut off at the scale discussed in the previous paragraph. For a voltage-biased tip, one should replace $I_0\to\omega{C}_\mathrm{tip}V_0$ in Eqs.~(\ref{Eq:P(r)},\ref{Eq:total_power}).

\section{Variational form of the heat transport equation}
\label{sec:functional}

\begin{figure}
\begin{center}
\includegraphics[width=0.45\textwidth]{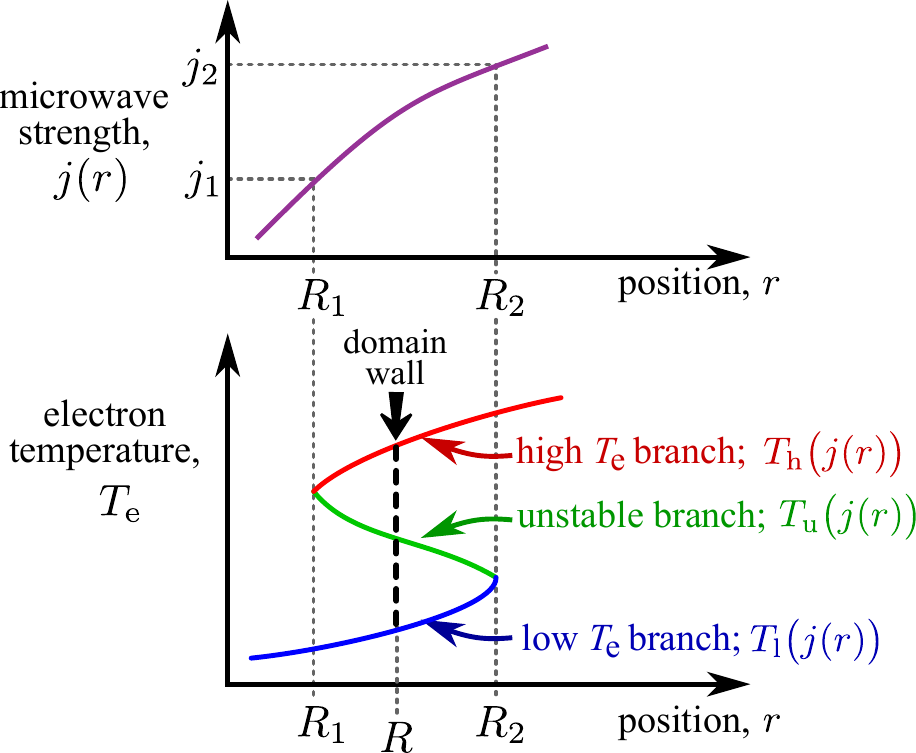}
\end{center}
\caption{
The situation for which Appendix~\ref{sec:functional} explains how to find the
domain wall's position, $R$. 
}
\label{fig_domainwall-construction}
\end{figure}

Eq.~(\ref{eq:HeatTransport}) can be written in terms of the variational derivative of a ``free energy'' functional~$\mathcal{F}[\te(\vec{r})]$:
\begin{subequations}\begin{align}
&\mathcal{K}(\te)\,c(\te)\,\frac{\partial\te}{\partial{t}}=-\frac{\delta\mathcal{F}[\te(\vec{r})]}{\delta\te(\vec{r})},\\
&\mathcal{F}=\int{d}^2\vec{r}\left[\frac{\mathcal{K}^2(\te)}2|\boldsymbol{\nabla}\te|^2
+\mathcal{Q}(\te)-j^2(r)\,\mathcal{R}(\te)\right],\label{freeenergy=}\\
&\mathcal{Q}(\te)=\int_0^{\te}\mathcal{K}(T')\,Q(T',T_\mathrm{ph})\,dT',\\
&\mathcal{R}(\te)=\int_0^{\te}\Re\frac{\mathcal{K}(T')}{\sigma(\omega,T')}\,dT'.
\end{align}\end{subequations}
A stable solution of Eq.~(\ref{eq:HeatTransport}) corresponds to a local minimum of the functional~(\ref{freeenergy=}).

We now use this to explain the domain wall's position,
taking the example of the situation sketched in Fig.~\ref{fig_domainwall-construction}.
There we assume  $\hbar\omega/T_c$ and $T_\mathrm{ph}/T_c$ are such that the system is bistable for $\mathcal{K}=0$ in the interval $j_1<j<j_2$ with some $j_1,j_2$.
In other words, for $j_1<j<j_2$, there are three $\te$ which are solutions of Eq.~(\ref{eq:HeatBalance}), these being 
$T_\mathrm{l}(j)<T_\mathrm{u}(j)<T_\mathrm{h}(j)$.
Let $R_1,R_2$ be such that $j(R_{1,2})=j_{1,2}$, and for definiteness we are assuming $R_1<R_2$.
If we take the limit $\Lambda/z_0\to0$ and neglect the gradient term, the position $R$ where the switching between the two solutions occurs can be found by minimizing the ``free energy''
\begin{align*}
&\int\limits_{R_1}^R2\pi{r}\,dr\left[j^2(r)\,\mathcal{R}(T_\mathrm{h}(j(r)))-\mathcal{Q}(T_\mathrm{h}(j(r)))\right]+{}\\
&{}+\int\limits_R^{R_2}2\pi{r}\,dr\left[j^2(r)\,\mathcal{R}(T_\mathrm{l}(j(r)))-\mathcal{Q}(T_\mathrm{l}(j(r)))\right]
\end{align*}
with respect to~$R$. This determines $R$ as the position where $j(R)$ satisfies the condition
\begin{align}
&\int_0^{T_\mathrm{l}(j(R))}\left[\Re\frac{j^2(R)\,\mathcal{K}(T)}{\sigma(\omega,T)}
-\mathcal{K}(T)\,Q(T,T_\mathrm{ph})\right]dT={}\nonumber\\
&{}=\int_0^{T_\mathrm{h}(j(R))}\left[\Re\frac{j^2(R)\,\mathcal{K}(T)}{\sigma(\omega,T)}
-\mathcal{K}(T)\,Q(T,T_\mathrm{ph})\right]dT,
\end{align}
which is equivalent to Eq.~(\ref{eq:domain_wall}). 

Including the weak gradient term, we can estimate the width~$w$ of the transition region by minimizing
\begin{align*}
&\min_w\left\{2\pi{R}w\left[\frac{\mathcal{K}(T_\mathrm{h})\,T_\mathrm{h}-\mathcal{K}(T_\mathrm{l})\,T_\mathrm{l}}{w}\right]^2\right.+{}\nonumber\\
&{}+\left.2\pi{R}w\int_{T_\mathrm{l}}^{T_\mathrm{u}}\left[\Re\frac{j^2\mathcal{K}(T)}{\sigma(\omega,T)}
-\mathcal{K}(T)\,Q(T,T_\mathrm{ph})\right]dT\right\}.
\end{align*}
If all temperatures are a fraction of~$T_c$, all quantities are of the same order as in the normal state. Then $j^2\sim\sigma{Q}$, which gives $w\sim\Lambda$.

\section{Heating by thermal radiation from a hot tip}
\label{app:thermal}

The body of the manuscript described in detail the use of the tip to apply near-field microwaves to the superconductor, however we only briefly mentioned heating the tip. Naively, one would expect the physics to be similar in both cases, because both processes are intended to locally heat the superconductor.  However, we find that heating the tip is a very ineffective way of heating the superconductor, compared to microwave driving.  For realistic experimental parameters, we find that a hot tip does not drive the superconductor into the normal state, nor create bistability. Here we explain in detail how we model the hot tip, and how we arrive at these conclusions.

\subsection{Hot tip as a fluctuating dipole}

As in Sec.~\ref{sec:microwave}, we model the superconducting layer as an infinitely thin 2D sheet with a 2D conductivity $\sigma_{2D}(\omega)=\sigma(\omega){d}$, and the tip as a small spherical metallic particle of radius~$R_\mathrm{tip}$, placed at a distance $z_0$ from the layer, assuming both ${d},R_\mathrm{tip}\ll{z}_0$.  The tip material is characterized by its bulk conductivity $\sigma_\mathrm{tip}$ which we assume to be frequency-independent.

A similar problem was studied in Ref.~\cite{Chapuis2008} in the framework of fluctuational electrodynamics~\cite{Rytov1953, Polder1971, Rytov1989}, which we will also adopt here. Namely, we represent the tip as a thermally fluctuating dipole (either electric or magnetic), which produces a fluctuating field which is heating up the electrons in the sheet. Below we analyze the electric and magnetic contributions separately, and find them to be of the same order. To simplify the calculations, we assume the dipole to oscillate only along the~$z$ direction (perpendicular to the sample plane). Contribution of the in-plane fluctuations is of the same order, so our results will remain valid as qualitative estimates.

In this appendix, since we are handling an essentially 3D problem, we adopt the notation $\vec{r}=(x,y,z)\equiv(\vec{r}_\|,z)$. In the rest of the paper, $\vec{r}$~refers to the in-plane position, that is, the subscript ``$\|$'' is omitted for brevity. For the in-plane wave vector we use the notation $\vec{k}\equiv(k_x,k_y)$. 

\subsection{Electric dipole}

According to the fluctuation-dissipation theorem, the electric dipole moment,
\begin{equation}
    p(t)=\int\frac{d\omega}{2\pi}\,p_\omega{e}^{-i\omega{t}},
\end{equation}
subject to thermal fluctuations at temperature $T_\mathrm{tip}$, has the fluctuation spectrum
\begin{equation}\label{eq:FDTelectric}
\langle{p}_\omega\,p_{\omega'}\rangle =
\hbar\Im\alpha_\mathrm{e}(\omega)\coth\frac{\hbar\omega}{2T_\mathrm{tip}}\,2\pi\delta(\omega+\omega'),
\end{equation}
where
\begin{equation}
    \alpha_\mathrm{e}(\omega) = 4\pi\varepsilon_0
    R_\mathrm{tip}^3\,\frac{i\sigma_\mathrm{tip}/(\varepsilon_0\omega)}{i\sigma_\mathrm{tip}/(\varepsilon_0\omega)+3}
\end{equation}
is the electric polarizability of a sphere of radius~$R_\mathrm{tip}$ with the dielectric function $1+i\sigma_\mathrm{tip}/(\varepsilon_0\omega)$. 
For a tip made of a good metal, $\sigma_\mathrm{tip}/\varepsilon_0\sim10^{18}\:\mbox{s}^{-1}$ (corresponding to $1/\sigma_\mathrm{tip}=1.13\times10^{-7}\:\Omega\cdot\mbox{m}$), we have $\omega\ll\sigma_\mathrm{tip}/\varepsilon_0$, so $\Im\alpha_\mathrm{e}\approx12\pi\varepsilon_0R_\mathrm{tip}^3\omega\varepsilon_0/\sigma_\mathrm{tip}$.

To find the induced fluctuating electric field, we use the quasistatic approximation, since the dimensions of the structure are much smaller than the thermal photon wavelength. Namely, we write the Poisson equation for each Fourier component of the electrostatic potential $\varphi_{\vec{k}\omega}(z)\,e^{i\vec{k}\vec{r}_\|-i\omega{t}}$:
\begin{align}
\left(\frac{\partial^2}{\partial z^2}-k^2\right)\varphi_{\vec{k}\omega}(z) =&-\frac{p_{\omega}}{\varepsilon_0}\,\delta'(z-z_0)\nonumber\\
&+\frac{ik^2}{\omega}\,\frac{\sigma_\mathrm{2D}(\omega)}{\varepsilon_0}\,\varphi_{\vec{k}\omega}(0)\,\delta(z).\label{eq:Poisson}
\end{align}
The charge density, appearing on the right-hand side, consists of two parts. The first one, $-p_{\omega}\delta'(z-z_0)$ with $\delta'(z)$ standing for the derivative of the Dirac delta function, is that of the point dipole $p_\omega$, placed at the point $\vec{r}=(0,0,z_0)$ and oriented along~$z$. The second part is the charge density induced in the superconducting layer at $z=0$ by the oscillating in-plane electric field $-i\vec{k}\varphi_{\vec{k}\omega}(0)$. Indeed, this field induces the current $-i\vec{k}\varphi_{\vec{k}\omega}(0)\,\sigma_\mathrm{2D}(\omega)$, which is related to the charge density by the continuity equation.

A solution of Eq.~(\ref{eq:Poisson}) is sought as linear combinations of $e^{\pm{k}z}$ in the three intervals $-\infty<z<0$, $0<z<z_0$, $z_0<z<\infty$, decaying at $\pm\infty$. The coefficients should be matched to give the correct jump of $\varphi_{\vec{k}\omega}(z)$ at $z=z_0$ and the jump in $d\varphi_{\vec{k}\omega}(z)/dz$ at $z=0$, needed to reproduce the right-hand side of Eq.~(\ref{eq:Poisson}). The result is
\begin{equation}\label{eq:phielectricdipole}
    \varphi_{\vec{k}\omega}(z=0) = \frac{p_\omega}{2\varepsilon_0}
    \frac{{e}^{-kz_0}}{1+k\,a_\mathrm{e}(\omega)}, \quad
a_\mathrm{e}(\omega)\equiv
\frac{i\sigma_\mathrm{2D}(\omega)}{2\varepsilon_0\omega}.
\end{equation}
Even for a rather small conductivity 
$\sigma_\mathrm{2D}=4e^2/(2\pi\hbar)=1/(6.46\:\mbox{k}\Omega)$
and a rather high frequency corresponding to the temperature $\hbar\omega=100\:\mbox{K}$, we obtain $|a_\mathrm{e}(\omega)|=670\:\mbox{nm}$, while the typical $k\sim1/z_0\sim(100\:\mbox{nm})^{-1}$. Thus, we can neglect unity in the denominator, and obtain the in-plane electric field $\vec{E}_{\|\omega}(\vec{r}_\|,z=0)$ by the inverse Fourier transform with respect to~$\vec{k}$:
\begin{equation}
 \vec{E}_{\|\omega}(\vec{r}_\|,z=0)= \frac1{4\pi\varepsilon_0}
 \frac{p_\omega}{a_\mathrm{e}(\omega)}\,
 \frac{\vec{r}_\|}{(r_\|^2+z_0^2)^{3/2}}.
\end{equation}
Finally, the heating power per unit area at a point $(\vec{r}_\|,0)$ is found as $\langle\vec{J}(\vec{r}_\|,t)\cdot\vec{E}_\|(\vec{r}_\|,0,t)\rangle$, where the current Fourier component $\vec{J}_\omega(\vec{r}_\|)=\sigma_{2D}(\omega)\,\vec{E}_{\|\omega}(\vec{r}_\|,0)$.  The averaging is performed using Eq.~(\ref{eq:FDTelectric}), and one should subtract the inverse heat flow from current fluctuations in the superconducting layer with electron temperature~$\te$, which is given by the same expression but with the replacement $\coth[\hbar\omega/(2T_{\mathrm{tip}})]\to\coth[\hbar\omega/(2\te)]$:
\begin{align}
P_\mathrm{e}(r_\|)=\frac{3}{\pi}\,\frac{r_\|^2R_\mathrm{tip}^3}{(r_\|^2+z_0^2)^3}
\int_{-\infty}^\infty\frac{d\omega}{2\pi}\,\hbar\omega^3\,\frac{\varepsilon_0}{\sigma_\mathrm{tip}}\,\Re\frac{\varepsilon_0}{\sigma_{2D}(\omega)}\nonumber\\
\times\left(\coth\frac{\hbar\omega}{2T_\mathrm{tip}}-\coth\frac{\hbar\omega}{2T_\mathrm{e}}\right).
\label{eq:Pe}
\end{align}

\subsection{Magnetic dipole}

The fluctuations of the magnetic moment~$m$ are fully analogous to Eq.~(\ref{eq:FDTelectric}):
\begin{equation}\label{eq:FDTmagnetic}
\langle{m}_\omega\,m_{\omega'}\rangle =\hbar\Im\alpha_\mathrm{m}(\omega)\coth\frac{\hbar\omega}{2T_\mathrm{tip}}\,2\pi\delta(\omega+\omega'),
\end{equation}
where
\begin{equation}
\alpha_\mathrm{m}(\omega)=\frac{2\pi}{15}\,i\omega\sigma_\mathrm{tip}R_\mathrm{tip}^5,
\end{equation}
is the magnetic polarizability of a sphere of radius~$R_\mathrm{tip}$ with the dielectric function $1+i\sigma_\mathrm{tip}/(\varepsilon_0\omega)$.
This expression can be obtained by calculating the magnetization corresponding to the circular currents produced by the electric field, which, in turn, is induced by the oscillating magnetic field, according to the Faraday's law.

In this magnetostatic problem, it is convenient to find the  electric field from the vector potential $\vec{A}_{\vec{k}\omega}(z)$. For the magnetic dipole $m\vec{e}_z$ directed along the $z$ axis with the unit vector $\vec{e}_z$, one can seek the vector potential in the form $\vec{A}_{\vec{k}\omega}(z)=i\vec{k}\times\vec{e}_z\,\psi_{\vec{k}\omega}(z)$. Substituting it into the 3D Amp\`ere's law $\boldsymbol{\nabla}\times\boldsymbol{\nabla}\times\vec{A}=\mu_0\vec{j}$ ($\mu_0=1/(\varepsilon_0c^2)$ being the vacuum magnetic permeability), we obtain the following equation for the scalar function $\psi_{\vec{k}\omega}(z)$:
\begin{align}
\left(\frac{\partial^2}{\partial z^2}-k^2\right)\psi_{\vec{k}\omega}(z)=&{}-\mu_0m_\omega\,\delta(z-z_0)\nonumber\\
&{}-i\omega\mu_0\,\sigma_\mathrm{2D}(\omega)\, \psi_{\vec{k}\omega}(0)\,\delta(z).
\end{align}
Proceeding analogously to the electrostatic case, we find the solution
\begin{align}
&\vec{A}_{\vec{k}\omega}(z=0)=\frac{\mu_0m_\omega}{2}\,
\frac{{e}^{-kz_0}}{k+1/a_\mathrm{m}(\omega)}\,i\vec{k}\times\vec{e}_z,\\
&a_\mathrm{m}(\omega)\equiv-\frac{2}{i\omega\mu_0\sigma_\mathrm{2D}(\omega)}.\nonumber
\end{align}
Again, taking $\sigma_\mathrm{2D}=4e^2/(2\pi\hbar)$ and $\hbar\omega=100\:\mbox{K}$, we obtain $|a_\mathrm{m}(\omega)|=0.8\:\mbox{mm}$, so  $1/a_\mathrm{m}$ in the denominator can be safely neglected even for much higher conductivities. The electric field,
\begin{equation}
 \vec{E}_{\|\omega}(\vec{r}_\|,z=0)=
 -\frac{i\omega\mu_0m_\omega}{4\pi}\,
 \frac{\vec{e}_z\times\vec{r}_\|}{(r_\|^2+z_0^2)^{3/2}},
\end{equation}
determines the heating power per unit area:
\begin{align}
P_\mathrm{m}(r_\|)=\frac{1}{120\pi}\frac{r_\|^2R_\mathrm{tip}^5}{(r_\|^2+z_0^2)^3}
\int_{-\infty}^\infty\frac{d\omega}{2\pi}\frac{\hbar\omega^3}{c^4}\,\frac{\sigma_\mathrm{tip}}{\varepsilon_0}\,\frac{\Re{\sigma_{2D}(\omega)}}{\varepsilon_0}\nonumber\\
\times\left(\coth\frac{\hbar\omega}{2T_\mathrm{tip}}-\coth\frac{\hbar\omega}{2T_\mathrm{e}}\right).
\label{eq:Pm}
\end{align}

%

\subsection{Electron temperature profile}

\begin{figure*}
\begin{center}
\includegraphics[width=0.45\textwidth]{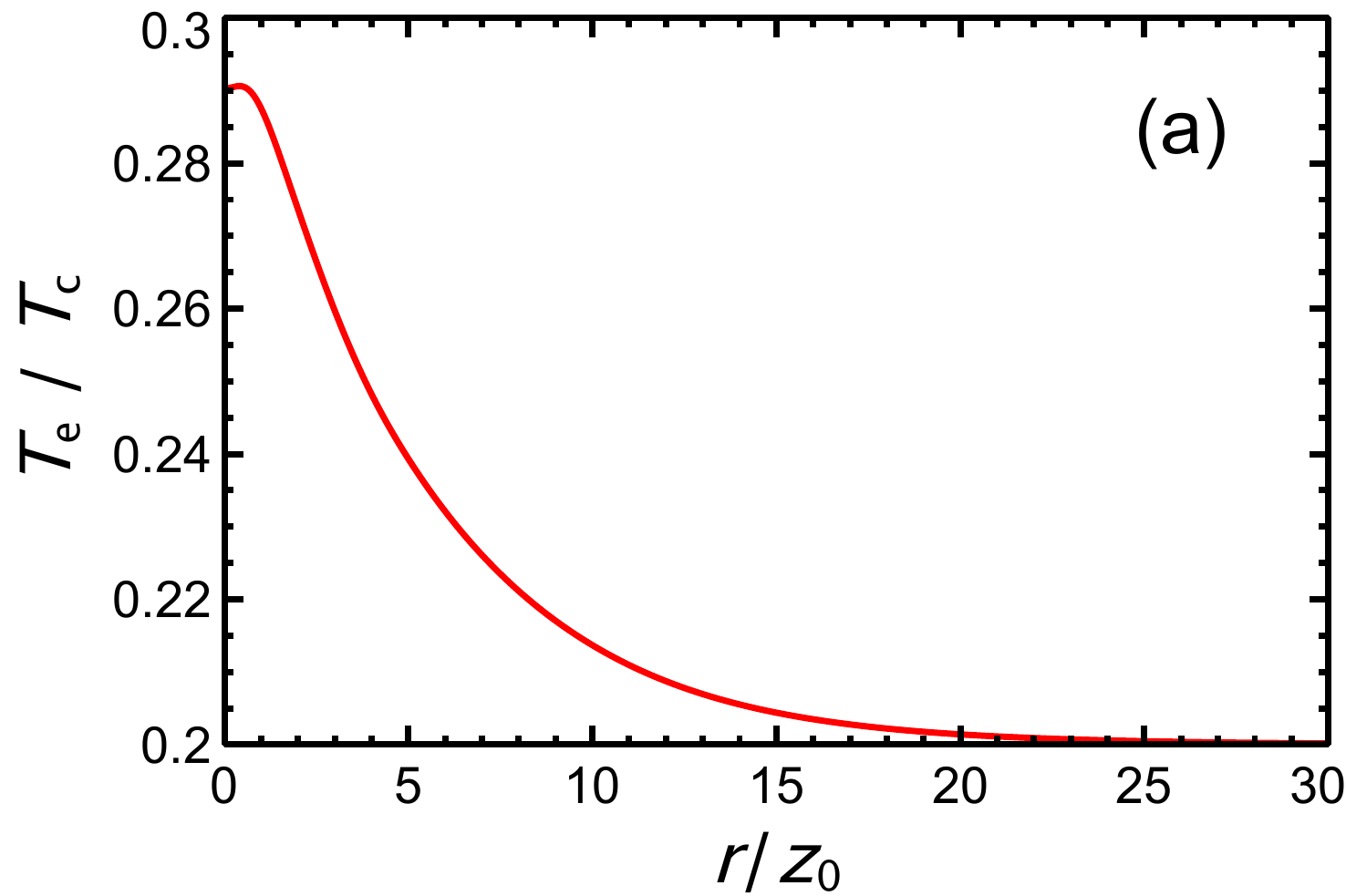}
\hfill
\includegraphics[width=0.45\textwidth]{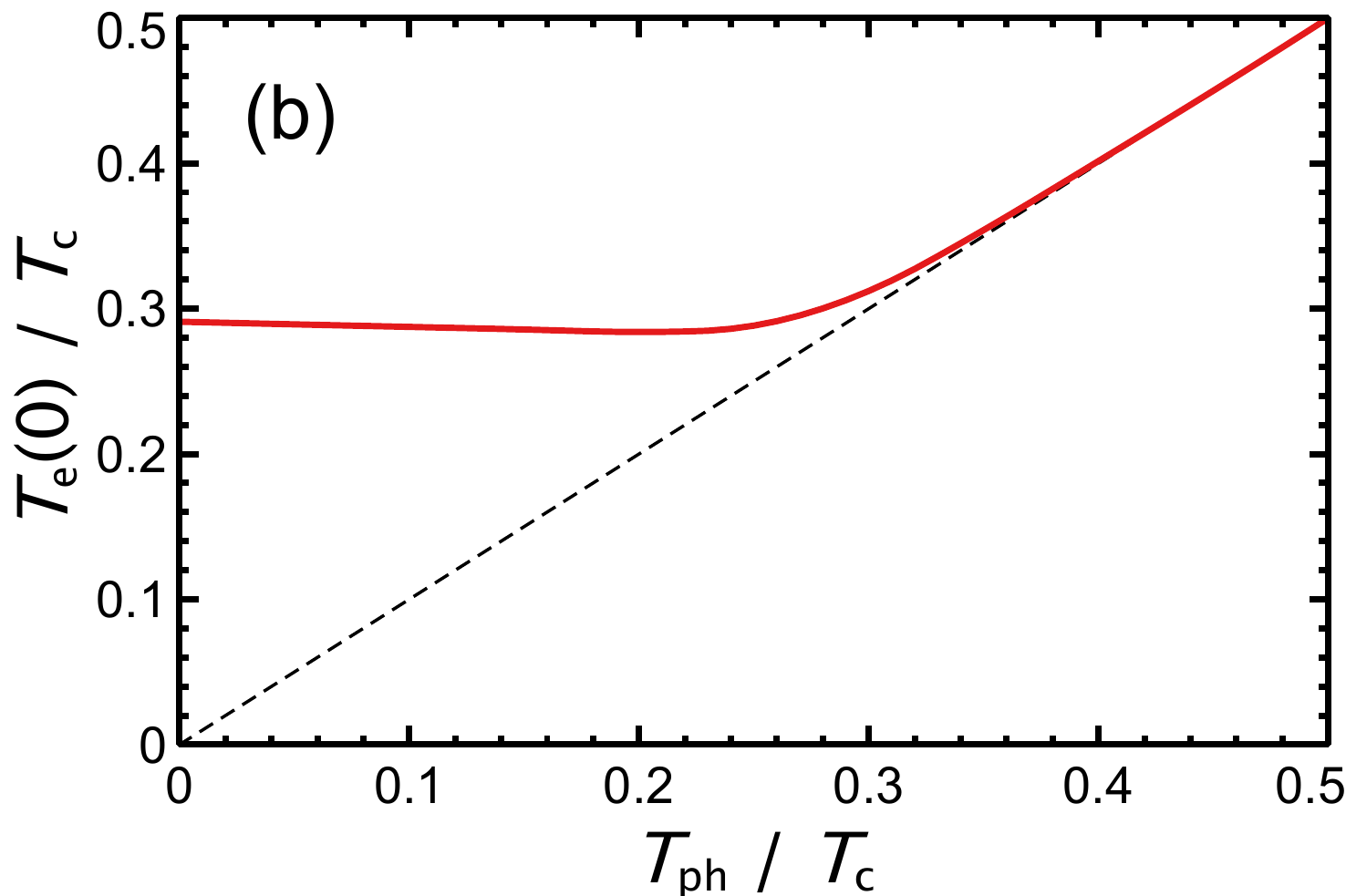}
\end{center}
\caption{Heating of the superconductor by a thermal tip for the parameters of InO$_x$ given in Table~\ref{tab:parameters},
$\sigma_\mathrm{tip}= 10^7\:(\Omega\cdot\mbox{m})^{-1}$, 
$R_\mathrm{tip}=50\:{\rm nm}$, $z_0=100\:{\rm nm}$, and $T_\mathrm{tip}=300\:\mbox{K}$.
(a)~Temperature profile $\te(r_\|)$ for $T_\mathrm{ph}/T_c=0.2$.
(b)~The dependence of $\te$ at the maximum ($r_\|=0$) on $T_\mathrm{ph}$ (solid curve). The dashed line is $\te=T_\mathrm{ph}$}
\label{fig_thermaltip}
\end{figure*}

As we will see shortly, the most interesting  case is $T_\mathrm{tip}\gg{T}_c$. Then one can neglect the frequency dependence of $\sigma(\omega)$ and take it to coincide with the normal state value. Then, the frequency integral in Eqs.~(\ref{eq:Pe}) and~(\ref{eq:Pm}) is calculated explicitly:
\begin{align}
P(r_\|)={}&{}\frac{\pi^2}{5}
\frac{T_\mathrm{tip}^4-\te^4}{\hbar^3c^2}
\frac{r_\|^2R_\mathrm{tip}^4}{(r_\|^2+z_0^2)^3}\nonumber\\
{}&{}\times\left(
\frac{c^2}{R_\mathrm{tip}}\,
\frac{\varepsilon_0}{\sigma_\mathrm{tip}}\,\frac{\varepsilon_0}{\sigma_{2D}}+
\frac1{360}\,\frac{R_\mathrm{tip}}{c^2}\,
\frac{\sigma_\mathrm{tip}}{\varepsilon_0}\,\frac{\sigma_{2D}}{\varepsilon_0}\right).\label{dee}
\end{align}
%
%
%
Let us estimate the heating power at $r_\|=z_0/\sqrt2$, corresponding to the maximum of $P(r_\|)$. For $z_0=100\:\mbox{nm}$, $R_\mathrm{tip}=50\:\mbox{nm}$,  $1/\sigma_\mathrm{tip}=10^{-7}\:\Omega\cdot\mbox{m}$, $\sigma_{2D}=4e^2/(2\pi\hbar)\approx1/(6.5\:\mbox{k}\Omega)$, and room temperature $T_\mathrm{tip}=300\:\mbox{K}$, we obtain $P\approx6\:\mbox{W}/\mbox{m}^2$ (the two terms in the brackets equal to 0.09 and 0.03, respectively, the electric dipole contribution being somewhat more important). Balancing it with the phonon cooling power $\Sigma(\te^n-T_\mathrm{ph}^n)d$, for the parameters of InO$_x$ (Table~\ref{tab:parameters} of the main text) and for $T_\mathrm{ph}=T_c=3\:\mbox{K}$, we obtain very little overheating, $\te-T_\mathrm{ph}\sim10^{-4}\:\mbox{K}$, and for NbN it is even smaller. A noticeable electronic overheating can be obtained for lower $T_\mathrm{ph}\lesssim1\:\mbox{K}$, when the cooling power is suppressed by the presence of the superconducting gap (Fig.~\ref{fig_thermaltip}). 
Still, it is not sufficient to suppress the superconductivity and create a normal region in the center.
Also, even for low $T_\mathrm{ph}$, the tip temperature $T_\mathrm{tip}$ should be quite high, so the heating occurs via absorption of photons with $\hbar\omega\sim{T}_\mathrm{tip}\gg{T}_c$. First, this validates Eq.~(\ref{dee}). Second, in this regime the heating power does not depend on~$\te$, which eliminates any bistability.

The typical size of the overheated region is at least~$z_0$. Moreover, at low temperatures the phonon cooling power $Q(\te,T_\mathrm{ph})$ vanishes faster than the electronic thermal conductivity $\mathcal{K}(\te)$. As a result, at temperatures significantly below $T_c$, the size of the overheated region significantly exceeds~$z_0$, which makes this setting unsuitable for creating a localized thermal perturbation.


%

\end{document}